\documentclass[%
 aip,
 amsmath,amssymb,
 reprint,%
]{revtex4-1}

\usepackage{graphicx}
\usepackage{dcolumn}
\usepackage{bm}

\usepackage[utf8]{inputenc}
\usepackage[T1]{fontenc}
\usepackage{mathptmx}
\usepackage{epsfig}
\usepackage{amsfonts}
\usepackage{mathrsfs}
\usepackage{amssymb,latexsym}

\newcommand{\fa}{\mathfrak{a}}

\newcommand{\fu}{\mathfrak{u}}

\newcommand{\fn}{{\,\mathfrak{n}\,}}

\newcommand{\fz}{\mathfrak{z}}

\newcommand{\fK}{\mathfrak{K}}

\newcommand{\bM}{\mathbf{M}}

\newcommand{\cJ}{\mathcal{J}}

\newcommand{\cP}{\mathcal{P}}

\newcommand{\cT}{\mathcal{T}}

\newcommand{\be}{\begin{equation}}
\newcommand{\ee}{\end{equation}}
\newcommand{\bea}{\begin{eqnarray}}
\newcommand{\eea}{\end{eqnarray}}
\newcommand{\nn}{\nonumber}

\newcommand{\ed}{\end{document}}

\newcommand{\ei}{\end{itemize}}

\newcommand{\bce}{\begin{center}}
\newcommand{\ece}{\end{center}}

\begin{document}

\preprint{AIP/123-QED}

\title[Broadband and Wide-Angle Invisibility with $\mathcal{P}\mathcal{T}$-Symmetric 2D-Weyl Semimetal]{Broadband and Wide-Angle Invisibility with $\mathcal{P}\mathcal{T}$-Symmetric 2D-Weyl Semimetal}

\author{Mustafa Sar{\i}saman}
 \email{mustafa.sarisaman@istanbul.edu.tr}
\affiliation{Department of Physics, Istanbul University, 34134,
Istanbul, Turkey
}%

\author{Murat~Tas}
\email{tasm236@gmail.com}
 \affiliation{%
Department of Basic Sciences, Alt{\i}nba\c{s} University, 34217 Istanbul, Turkey
}%

\date{\today}

\begin{abstract}
Inspired by the magnificent features of two-dimensional (2D) materials which aroused much of the interest
in recent materials science research, we study $\cP\cT$-symmetric 2D Weyl semimetal (WSM) to reveal the
broadband and wide-angle invisible configurations in a $\cP\cT$-symmetric optical slab system. Desired
unidirectional reflectionlessness and invisibility phenomena is obtained by the optimal control of system
parameters. We unravel the mystery of broadband and wide-angle invisibility in regular slab materials
with finite refractive indices by means of the plenary expressions. We show that materials whose
refractive indices relatively small (usually around $\eta =1$) give rise to quite a lot broadband and
wide-angle (almost all incidence angles) invisible configurations. This is not observed with any 2D
material other than 2D WSMs. Our findings suggest a concrete expedience to experimental realizations in
this direction.
\end{abstract}

\maketitle

\section{Introduction}

$\mathcal{PT}$ symmetry prodigy~\cite{bender-1, bender-2} has raised a remarkable development in quantum mechanics
since its early discovery. In particular, its practicable scopes in the fields of quantum optics and
condensed matter physics have attracted most of the recent interest in the last decade, and thus it has
led to advance numerous studies and applications~\cite{ijgmmp-2010-1,ijgmmp-2010-2,bagchi,PT1,PT2-1,PT2-2,PT4-1,PT4-2,PT6,PT7,PT8,PT9-1,PT9-2,PT9-3,PTnew1,PTnew2,PTnew3,PTnew4}.
It has been shown that the advantage of placing gain and loss layers side by side~\cite{Regensburger}
inspired several intriguing optical phenomena and devices, such as dynamic power oscillations of light
propagation, coherent perfect absorber lasers~\cite{CPA-1,CPA-2,CPA-3,CPA-4,CPA-5,lastpaper}, spectral singularities~\cite{naimark-1,naimark-2,
p123,pra-2012a,longhi4,longhi3} and unidirectional invisibility~\cite{PT1,PT6,PT7,pra-2017a,
invisibilitypapers-1,invisibilitypapers-2,invisibilitypapers-3,invisibilitypapers-4,invisibilitypapers-5,invisibilitypapers-6,invisibilitypapers-7,longhi1,pra-2013a,pra-2014b,midya,soto,lin1,pra-2015b,pra-2015a}. Strictly speaking, a Hamiltonian endowed with $\cP\cT$-symmetry contains a potential whose feature is displayed by
$V(x) = V^{\ast}(-x)$~\cite{bender-1,bender-2,PT1,PT4-1,PT4-2}. In general, complex $\cP\cT$-symmetric optical potentials
can be achieved by means of complex refractive indices, such that their optical modulations in complex
dielectric permittivity plane results in both optical absorption and amplification.

\begin{figure}
\begin{center}
\includegraphics[scale=.40]{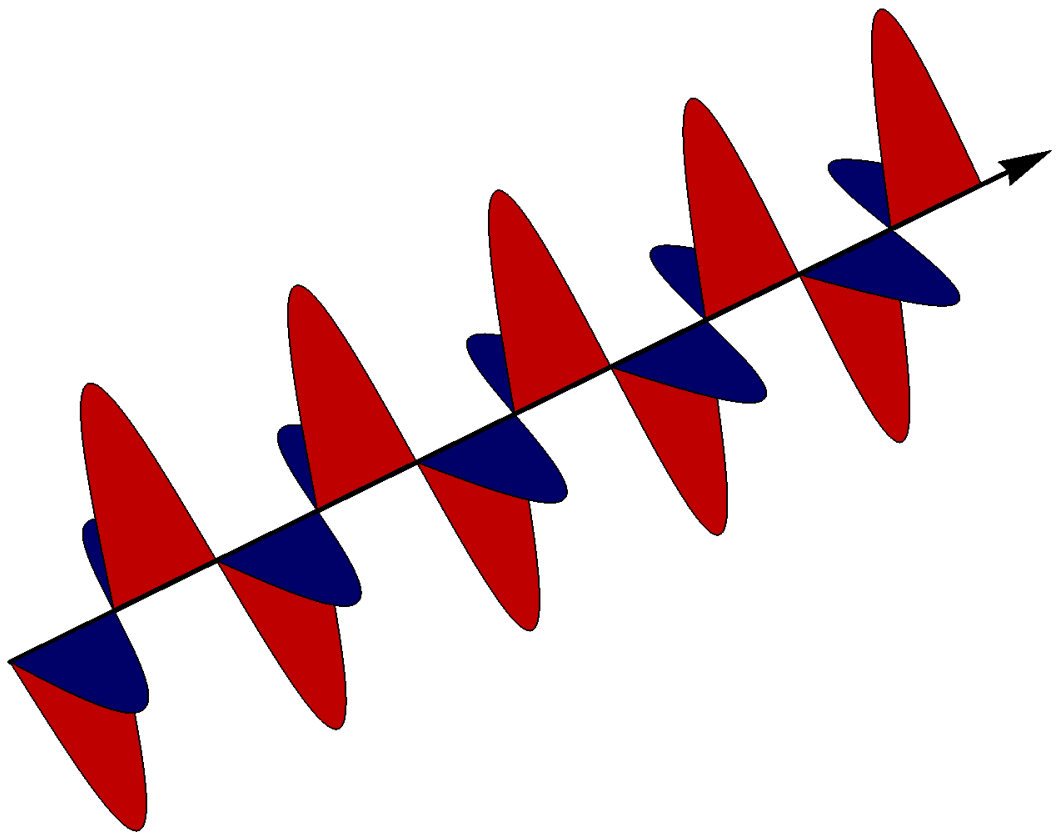}
\includegraphics[scale=.40]{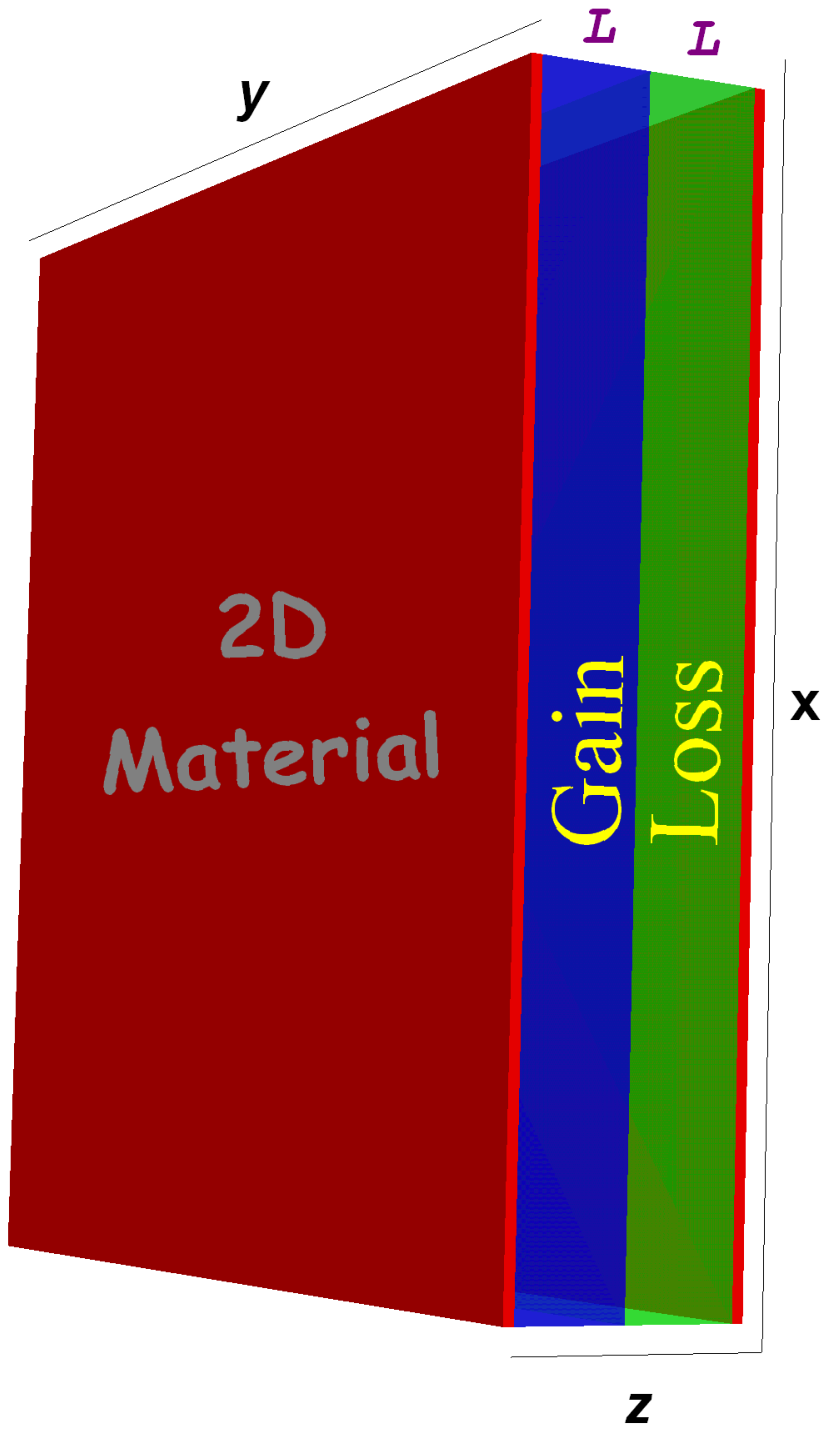}
\caption{(Color online) TE mode configuration for the parallel pair of optically active system covered by
the 2D material sheets obeying the $\cP\cT$-symmetry. Wave is emitted on the slab system by an angle
$\theta$ which is measured from the normal to the surface of 2D material.}
\label{fig1}
\end{center}
\end{figure}

Invisibility studies essentially initiated about a decade ago by two separate, but concurrent novel papers
by Pendry and Leonhardt~\cite{pendry-1,pendry-2}, which employ the transformation optics approach. Now, there is a
huge body of literature utilizing this method on the invisibility cloaking in photonic
systems~\cite{invpapers-1,invpapers-2,invpapers-3,invpapers-4,invpapers-5,invpapers-6,invpapers-7,invpapers-10,invpapers-11,invpapers-12,invpapers-13,invpapers-15,invpapers-16,invpapers-17}. Another approach in this field employs the transfer matrix formalism, which is
our point of interest in this study. It is revealed that both gain and loss layers are required for an
invisible configuration. In this respect, $\cP\cT$-symmetry helps the related parameters of the system be
adjusted in order to realize the invisibility. It is noted that a linear homogeneous slab material
equipped with $\cP\cT$-symmetry can not produce a broadband and wide angle invisibility~\cite{pra-2017a,sarisaman2019a},
i.e. the extend of invisibility remains in rather restricted spectral range and incidence angle. The use
of graphene sorts out this problem to some extend~\cite{cpa3, sarisaman2019b}, although spectral range could be increased
considerably, the range of incidence angle could not be widened commonly. Therefore, in view of this
motivation, we aim to employ 2D Weyl semimetal (WSM) to settle up this problem, and achieve broadband and
wide-angle invisibility phenomenon of a $\cP\cT$-symmetric optically active slab system covered by
$\cP\cT$-symmetric 2D WSM, see Fig.~\ref{fig1}.

Recent advances on the 2D materials have triggered development of a vast area of research, not only about
their physical properties, but also their applications in various fields, especially in
optics~\cite{gr5-1,gr5-2,gr5-3,gr5-4,gr5-5,gr5-6,gr5-7,gr6,gr7,gr8-1,gr8-2,naserpour,sarisaman2019c}. Among all of the 2D materials, graphene has the pioneering role
due to its well-documented physical properties and numerous applications~\cite{gr1-1,gr1-2,gr1-3,gr1-4,2d1}. As the family of
2D materials expands to include new members such as 2D WSMs, 2D semiconductors, boron nitride and more
recently, transition metal dichalcogenides and Xenes, atomically thin forms of these materials offer
endless possibilities for fundamental research, as well as demonstration of improved or even entirely
novel technologies~\cite{2d1,2d2-1,2d2-2,2d2-3}. New 2D material treatments could unleash new uses. Exciting properties
of 2D materials reveal that they may interact with electromagnetic waves in anomalous and exotic ways,
providing new phenomena and applications. Thus, new distinctive studies of invisibility phenomena with
2D materials have arisen. Especially recent works in this field originate essential motivation of our
work~\cite{lastpaper,graphene-1,graphene-2,graphene-3,graphene-4,graphene-5,prl-2009}. In this study, we offer one of the potential applications of 2D
materials in the invisibility phenomenon together with $\cP\cT$-symmetry attribute in optical systems.

In our analysis, we present all possible configurations of solutions that support the broadband and
wide-angle reflectionlessness and invisibility, schematically demonstrate their behaviors and show the
effects of various parameters. In particular, we demonstrate that optimal control of the slab and 2D
material parameters, i.e., gain coefficient, incidence angle, slab thickness, and Weyl nodes separation
$b$ give rise to broadband and wide-angle invisibility. We present exact conditions and display the role
of $\cP\cT$-symmetric 2D material in the gain adjustment. Our findings can be experimentally realized
with ease. Our method and hence results are quite reliable for all realistic slab materials of practical
concern.

\section{TE Mode Solution and Transfer Matrix}\label{S2}

Consider a parallel pair of optically active gain and loss slab system covered by the same kind of 2D
material at both ends which are enforced to respect the overall $\cP\cT$-symmetry, as depicted in
Fig.~\ref{fig1}. The gain and loss regions of slab have the same thickness $L$, and are restricted to
the regions $0<z<L$ and $L<z<2L$ with refractive indices $\fn_1$ and $\fn_2$ respectively. Assuming
time-harmonic electromagnetic waves are hitting on the slab, Helmholtz equation associated with TE mode
states is expressed as
    \begin{eqnarray}
    &\left[\nabla^{2} + k^2\fz_{j}(z)\right] \vec{E}^{j} = 0, \qquad
    \vec{H}^{j} = -\frac{i}{k Z_{0}} \vec{\nabla} \times \vec{E}^{j},
    \label{equation4}
    \end{eqnarray}
where $k:=\omega/c$ is the wavenumber, $c:=1/\sqrt{\mu_{0}\varepsilon_{0}}$ is the speed of light in
vacuum, and $Z_{0}:=\sqrt{\mu_{0}/\varepsilon_{0}}$ is the impedance of the vacuum. The index $j$
represents the regions of space respectively corresponding to $z<0$, $0<z<L$, $L<z<2L$ and $z>2L$,
thus $j = 0, 1, 2, 3$. The complex quantity $\fz_{j}(z)$ is
    \begin{eqnarray}
    \fz_{j}(z):= \fn_j^{2} ~~~~~ {\rm for~}z\in z_{j}.
    \label{e1}
    \end{eqnarray}
Notice that $\fn_0 = \fn_3 = 1$ in vacuum, and hence $\fz_{j}(z) = 1$ outside the slab. Therefore, TE
wave solution of (\ref{equation4}) is obtained as
    \begin{eqnarray}
    \vec E^{j}(\vec{r}) = \mathcal{E}^{j}(z) e^{i k_x x} \hat{e_y},
    \label{ez1}
    \end{eqnarray}
where $\hat e_y$ denotes the unit vector in $y$-direction, and $k_x$ and $k_z$ are the components of
wavevector $\vec{k}$ in $x-z$ plane. Thus, it is immediate to attain $\mathcal{E}^{j}$ by use of the
Helmholtz equation in (\ref{equation4}) as
    \be
   \mathcal{E}^{j} (L \mathbf{z})=a_j\,e^{i{\fK}_j \mathbf{z}} + b_j\,e^{-i{\fK}_j \mathbf{z}} ~~~~~~
   {\rm for} ~~ \mathbf{z}\in \mathbf{z}_{j}.
    \label{E-theta}
    \ee
Here $a_j$ and $b_j$ are possibly $k$-dependent amplitudes, and scaled variables $\mathbf{z}$ and $\fK$
are defined as
    \begin{eqnarray}
    \mathbf{z}:=z/L,   \qquad \fK:=Lk_z=kL\cos\theta . \label{scaled-var}
    \end{eqnarray}
Hence ${\fK}_j$ is specified by
    \begin{eqnarray}
    {\fK}_j:=\fK \tilde\fn_j,  \qquad \tilde\fn_j := \sec\theta \sqrt{\fn^2_j -\sin^2\theta}.
    \label{tilde-parm}
    \end{eqnarray}

Standard boundary conditions at $\mathbf{z} = 0, 1, 2$ yield coefficients $a_j$ and $b_j$, which include
substantial information about the 2D material at the slab ends, and is manifested by the conductivity of
the sheets. The conductivity takes a tensorial form and results in the surface current
$\vec{\cJ}^{(\ell)}_j := \sigma^{(\ell)}_j\, \vec{E}^j$, where $\sigma_j$ is the conductivity tensor at
$j$-th surface, and $\ell$ denotes the 2D material type. Our general formalism below is well-established
for all 2D materials with only diagonal or off-diagonal conductivity tensor, and hence
$\sigma^{(\ell)}_j$ can be regarded as scalar. Here we will specifically consider the 2D WSMs.

We emphasize that the boundary conditions associate the coefficients $a_j$ and $b_j$, and help to built
the transfer matrix which leads to extracting information about the reflection and transmission
coefficients of the optical system. Waves on the right-hand side are related to the waves on the
left-hand side by the transfer matrix $\bM$ as
     \begin{eqnarray}
    \left[\begin{array}{c}
    a_3\\ b_3\end{array}\right]=\bM \left[\begin{array}{c}
    a_0\\ b_0\end{array}\right].
    \label{transfermatrix}
    \end{eqnarray}
Hence, components of the transfer matrix in (\ref{transfermatrix}) can be obtained as follows
    \bea
    M_{\alpha\beta} &=\fa_{\alpha}\left[U_{\beta}(\fu_{\alpha}^{(2)}+\alpha)e^{i{\fK}_2} +
    V_{\beta}(\fu_{\alpha}^{(2)}-\alpha)e^{-i{\fK}_2}\right],
    \label{M22=x}
    \eea
where subindices $\alpha, \beta$ denote $+, -$, and $M_{++}$ represents the $M_{11}$ component and so
forth. Also
    \begin{eqnarray*}
    \fa_{\alpha} := \frac{e^{-2i\alpha\fK}}{8},~~~~~~~~~~~~
    \fu_{\alpha}^{(j)} := \frac{1+\alpha\,\sigma_j^{(\ell)}}{\tilde\fn_{j}},\nn\\
    U_{\alpha} := \sum_{\gamma = -1}^{+1} (\tilde{\fn}_2 +
    \gamma\,\tilde{\fn}_1) (1 + \alpha\,\gamma\,\fu_{\alpha}^{(1)})e^{i\gamma{\fK}_1},\nn\\
    V_{\alpha} := \sum_{\gamma = -1}^{+1} (\tilde{\fn}_2 -
    \gamma\,\tilde{\fn}_1) (1 + \alpha\,\gamma\,\fu_{\alpha}^{(1)})e^{i\gamma{\fK}_1},
    \nn
    \end{eqnarray*}
with $j = 1, 2$. Since the slab system is overall $\mathcal{PT}$-symmetric, components of $\bM$ satisfy the
relations in~\cite{pra-2014c}
    \begin{eqnarray}
    M_{\alpha\beta}\stackrel{\mathcal{PT}}{\longleftrightarrow}M^{*}_{\beta\alpha},
    \qquad M_{\alpha\beta}\stackrel{\mathcal{PT}}{\longleftrightarrow}-M^{*}_{\alpha\beta}.\nn
    \end{eqnarray}
Notice that $i\alpha$ and $\alpha\,\sigma_j^{(\ell)}$ are invariant under $\mathcal{PT}$-symmetry. One thus
obtains the following $\mathcal{PT}$ symmetric relations
    \begin{eqnarray}
    \fn_1\stackrel{ \mathcal{PT} }{\longleftrightarrow}\fn_2,
    \qquad \tilde{\fn}_1\stackrel{ \mathcal{PT} }{\longleftrightarrow}\tilde{\fn}_2,
    \qquad \fa_{+}\stackrel{\mathcal{PT}}{\longleftrightarrow}\fa_{-},\nn\\
    \fu_{\pm}^{(1)}\stackrel{ \mathcal{PT} }{\longleftrightarrow}\fu_{\mp}^{(2)},
    \qquad \sigma_1^{(\ell)}\stackrel{ \mathcal{PT} }{\longleftrightarrow}-\sigma_2^{(\ell)}.
    \label{pt-symmetry-rels}
    \end{eqnarray}
The last relation reveals that currents on the 2D material sheets flow in opposite directions. Hence, the
left and right reflection and transmission coefficients are obtained by means of the transfer matrix as
   \be
   \textrm{R}^{l} = -\zeta_{+}\chi^{-1},~~~\textrm{R}^{r} = \zeta_{-}\chi^{-1}\,e^{-4i{\fK}},~~~
   \textrm{T} =\fa_{-}^{-1} \chi^{-1}, \label{leftreflcoef}
   \ee
where
\bea
\zeta_{\alpha} &:= U_{\alpha}(\fu_{-\alpha}^{(2)} - \alpha)e^{i{\fK}_2} +
V_{\alpha}(\fu_{-\alpha}^{(2)} + \alpha)e^{-i{\fK}_2}, \label{zeta}\\
\chi &:= U_{-}(\fu_{-}^{(2)}-1)e^{i{\fK}_2} + V_{-}(\fu_{-}^{(2)}+1)e^{-i{\fK}_2}.\label{chi}
\eea
Notice that these expressions are applicable to all 2D materials having the scalar conductivity
$\sigma_j^{(\ell)}$. In this study, we consider 2D WSM \footnote{The case of graphene was studied in
\cite{cpa3} comprehensively. Here, we examine the 2D WSM in detail, and compare it to the graphene.} as
2D material which has conductivity $\sigma_j^{(w)}$. For 2D WSM, the conductivity of each sheet is
computed by using the Kubo formalism \cite{weyl-conductivity} as
   \begin{eqnarray}
   \sigma^{(w)} \approx i \int_{\xi < \lambda} \frac{dk_z}{2\pi} \sigma^{2D} (k_z) \xi (k_z)
   \simeq \frac{i e^2}{\pi h} \ln (2b\lambda),  \label{conductivityweyl}
   \end{eqnarray}
where surface state labeled by $k_z$ is localized near the sheets with a localization length
$\xi (k_z) = 2b/(b^2 -k_z^2)$, $\sigma^{2D} (k_z)$ is 2D quantized Hall conductivity, and
$\sigma^{2D} (k_z) = e^2/h$, and $b$ is the measure of separation between Weyl nodes given by
$b=\textbf{b} /\hat{e_z}$ in $k_z$-space. Here the symbol '$\approx$' is used to imply that real part of $\sigma^{(w)}$ is negligibly small compared to the imaginary part. Therefore, we say that $\sigma^{(w)}$ is pure imaginary as distinct from
the case of graphene \cite{cpa3}.

We emphasize that the left/right reflection and transmission coefficients contribute the necessary
information about the unidirectional reflectionless and invisible configurations of the optical system,
which we explore next.

\section{Invisibility Conditions and Related Parameters}\label{S4}

An optical system is designed to be left/right reflectionless if $\textrm{R}^{l/r} = 0$ together with
$\textrm{R}^{r/l} \neq 0$ is satisfied simultaneously. Moreover, one must provide the condition
$\textrm{T}=1$ to realize the left/right invisibility. Unidirectionally reflectionless condition is
achieved once both of the following conditions are satisfied
    \begin{eqnarray}
    e^{-2i{\fK}_2} = \mathcal{F}_{\alpha},~~~~~ e^{-2i{\fK}_2} \neq \mathcal{F}_{-\alpha}.
    \label{unidir-refl}
    \end{eqnarray}
Here $\mathcal{F}_{\alpha}:= \left[U_{\alpha} \left(1 - \alpha \fu_{-\alpha}^{(2)}\right)\right] /
\left[V_{\alpha} \left(1 + \alpha \fu_{-\alpha}^{(2)}\right)\right]$, and equations with $\alpha = +$
and $-$ correspond to the left and right reflectionlessness respectively. Unidirectional invisibility
exists when the condition $\textrm{T}=1$ is imposed in addition to (\ref{unidir-refl}), which yields the
condition for left/right invisibility as follows
   \begin{eqnarray}
   e^{- 4 i \alpha {\fK}} = \mathcal{G}_{\alpha}, ~~~~~ e^{- 4 i \alpha {\fK}} \neq \mathcal{G}_{-\alpha}.
   \label{invsibilitycondition1}
   \end{eqnarray}
where $\mathcal{G}_{\alpha} := 4\tilde{\fn}_2^2 \left(1-\left[\fu_{-\alpha}^{(2)}\right]^2\right) /
\left(U_{\alpha}V_{\alpha}\right)$. Eqs.~\ref{unidir-refl} and \ref{invsibilitycondition1} are in fact
complex expressions displaying the behavior of system parameters leading to the unidirectional
reflectionlessness and invisibility conditions respectively. Thus, thorough physical consequences of
these expressions can be examined by an in-depth analysis. First of all, by virtue of $\mathcal{PT}$-symmetry
relations in (\ref{pt-symmetry-rels}), we address the following specifications
    \begin{eqnarray}
    \fn:= \fn_1 =\fn_2^*,
    \qquad \sigma^{(\ell)}:= \sigma^{(\ell)}_{1}= -\sigma^{(\ell)\ast}_{2}.
    \label{eq251}
    \end{eqnarray}
We express the real and imaginary parts of refractive index $\fn$ as $\fn = \eta + i \kappa$ such that
the condition $\left|\kappa\right| \ll \eta - 1 < \eta$ holds by most of the materials. Finally, we
introduce another significant parameter, the gain coefficient $g$ as
   \begin{eqnarray}
   g:=-2k\kappa = -\frac{4\pi\kappa}{\lambda}.\label{gaincoef}
   \end{eqnarray}
We replace quantities $\sigma^{(\ell)}$, $\fn$ and $g$ into the Eqs.~\ref{unidir-refl} and
\ref{invsibilitycondition1} in order to obtain the reflectionless and invisible behaviors of our optical
system. The presence of 2D materials is prevalent explicitly with the term $\fu_{\alpha}^{(j)}$ which
implicitly contains $\sigma^{(\ell)}$. Thus, desired optimal conditions appear by the selection of most
appropriate system parameters. We perform a graphical analysis to seek out the act of each parameter by
means of various plots of the gain coefficient $g$. We take our slab to consist of Nd:YAG
crystals~\footnote{We do not consider the presence of dispersion. Resonance occurs at wavelength
$\lambda_0 = 808~\textrm{nm}$ for Nd:YAG.} with $\eta = 1.8217$, slab thickness $L=1~\textrm{cm}$ and
incidence angle $\theta = 30^{\circ}$~\cite{silfvast}. For the 2D WSM, we employ the parameters of
$b = 0. 05$ {\AA} and wavelength $\lambda = 808~\textrm{nm}$ which correspond to the resonance of Nd:YAG
crystals \footnote{Here and in what follows we disregard the experimental difficulties of maintaining
equal amounts of gain and loss in the relevant components of our system using Nd:YAG crystals coated by
equal amounts of 2D WSM. We employ numerical values associated with this system to demonstrate general
behaviors of the model we consider.}. Other parameters are displayed in the figures.

\begin{figure}
    \begin{center}
    \includegraphics[scale=.45]{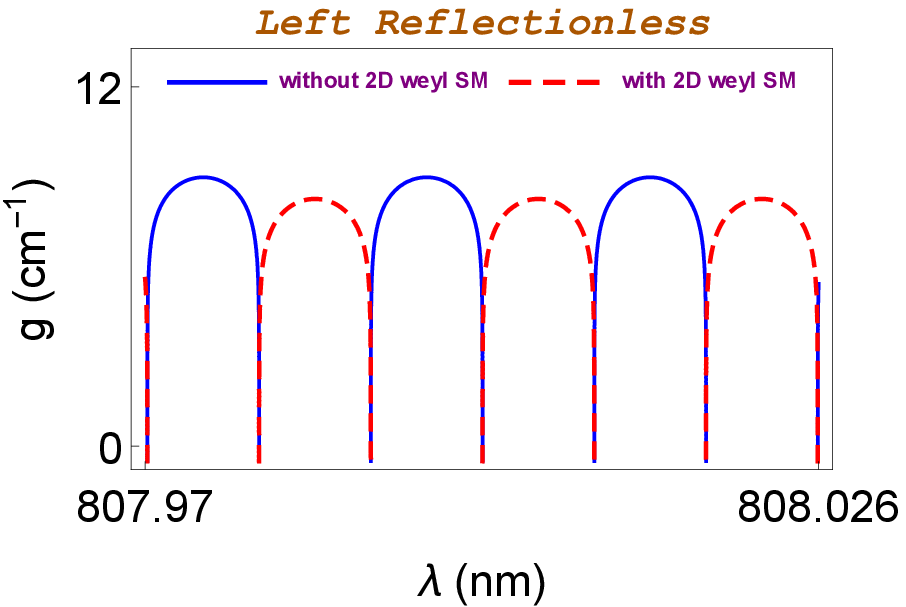}~~~
    \includegraphics[scale=.45]{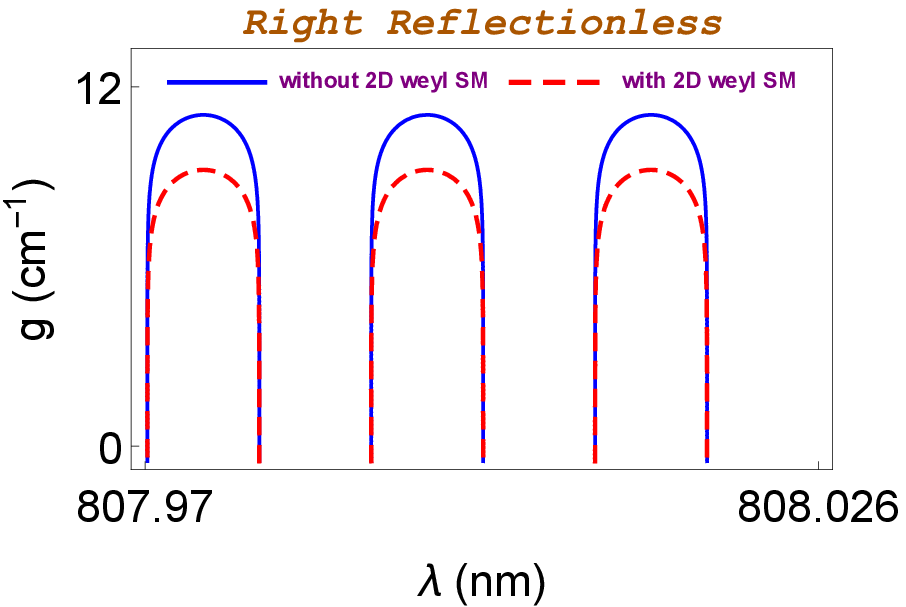}
    \caption{(Color online) Behavior of the gain $g$ as a function of wavelength $\lambda$ corresponding
to the cases of with and without 2D Weyl SM.}
    \label{fig1m}
    \end{center}
    \end{figure}

Figure~\ref{fig1m} exhibits the behavior of gain coefficient in response to the wavelength $\lambda$ of
incoming waves for the left and right reflectionless cases. We observe that whereas both left and right
reflectionless cases in the absence of 2D WSM happen to exist at the same wavelength ranges which form
periodically repeated patterns, this turns into a new configuration in the presence of 2D WSM at which
left reflectionless wavelength ranges totally shift to the wavelength ranges previously yielding no
reflectionlessness while right reflectionless wavelength ranges remain the same. This explicitly shows
that placing $\mathcal{PT}$-symmetric 2D WSM around a $\mathcal{PT}$-symmetric slab is highly effective in obtaining
a drastic change in the gain value. We can not simultaneously observe left and right reflectionlessness
in the presence of 2D WSM except for bidirectional reflectionlessness. We also notice that the gain value
in case of uni- or bidirectionally reflectionless cases reduces subject to the incident wavelength
$\lambda$ and $b$-value. We reveal that this engenders a rather intriguing situation that is completely
different from graphene. In the case of graphene, specified wavelength range never changes while the gain
value reduces similarly to 2D WSM. This causes 2D WSM to show unexpected fascinating features in respect
of unidirectional reflectionlessness and invisibility.

\begin{figure}
    \begin{center}
    \includegraphics[scale=.45]{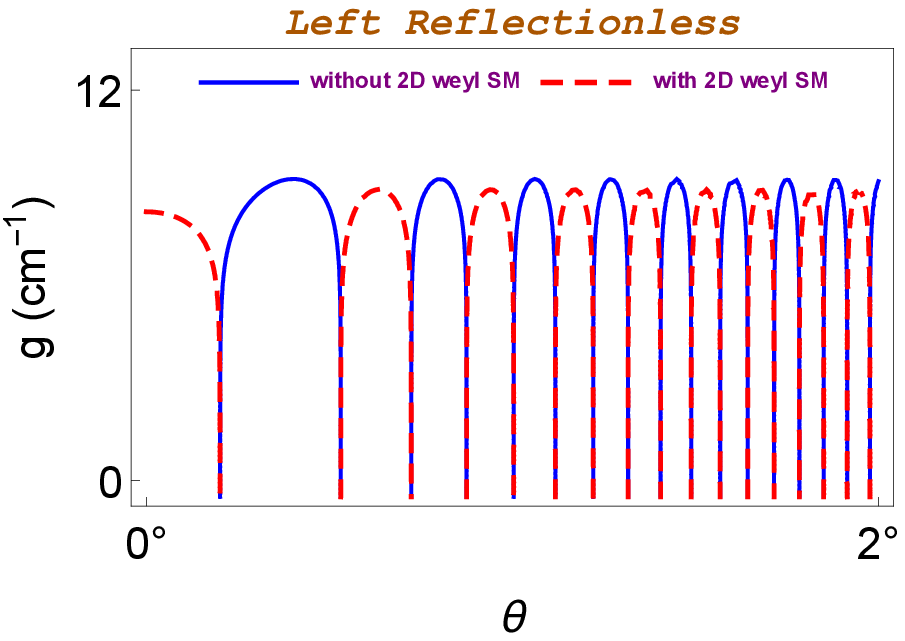}~~~
    \includegraphics[scale=.45]{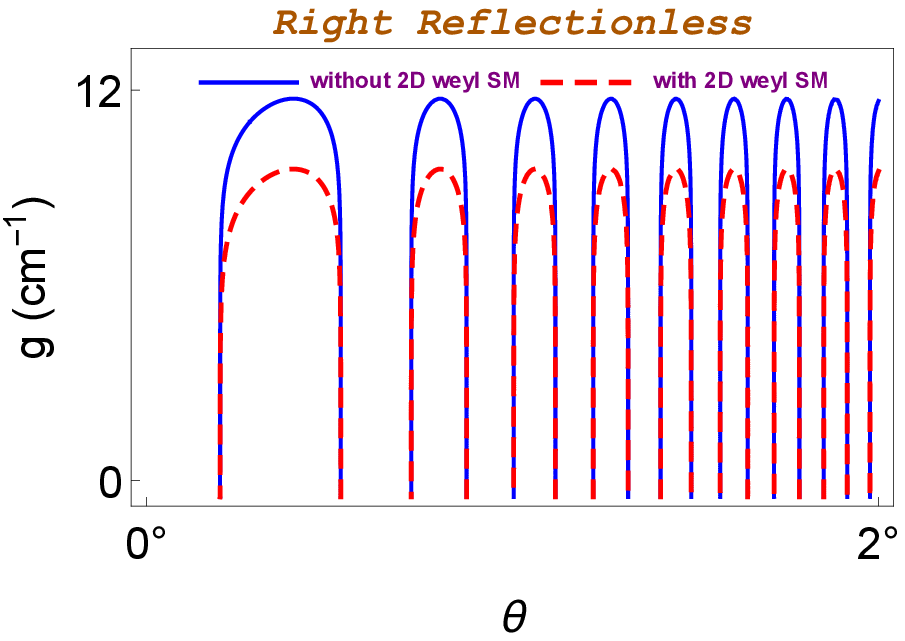}
    \caption{(Color online) The gain value as a function of incidence angle $\theta$, which yields left
and right reflectionless configurations.}
    \label{fig3m}
    \end{center}
    \end{figure}

A similar behavior is observed in incidence angle vs. gain coefficient plot depicted in Fig.~\ref{fig3m}.
Here we adapt wavelength $\lambda = 808~\textrm{nm}$. Again when the 2D WSM sheets are placed at the ends
of the slab, we encounter the left reflectionless patterns which are fully shifted in the angle ranges
that no left reflectionlessness is observed in the absence of 2D WSM. Right reflectionless case yields
similar behavior as in Fig.~\ref{fig1m} Indeed, we notice that left reflectionlessness occurs at
incidence angle $\theta = 0$, which does not appear without the 2D WSM. Corresponding to the same angle
ranges, only a certain direction of reflectionlessness is observed.

\begin{figure}
    \begin{center}
    \includegraphics[scale=.32]{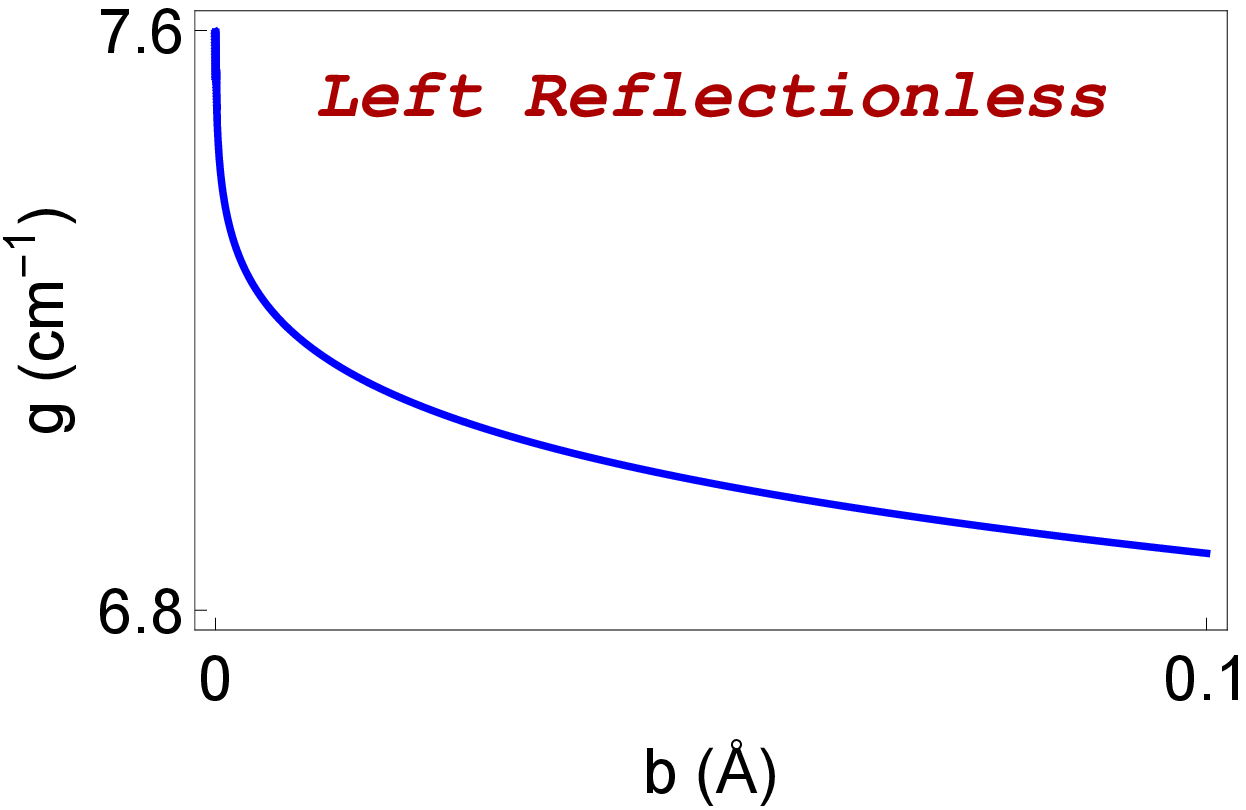}~~~
    \includegraphics[scale=.32]{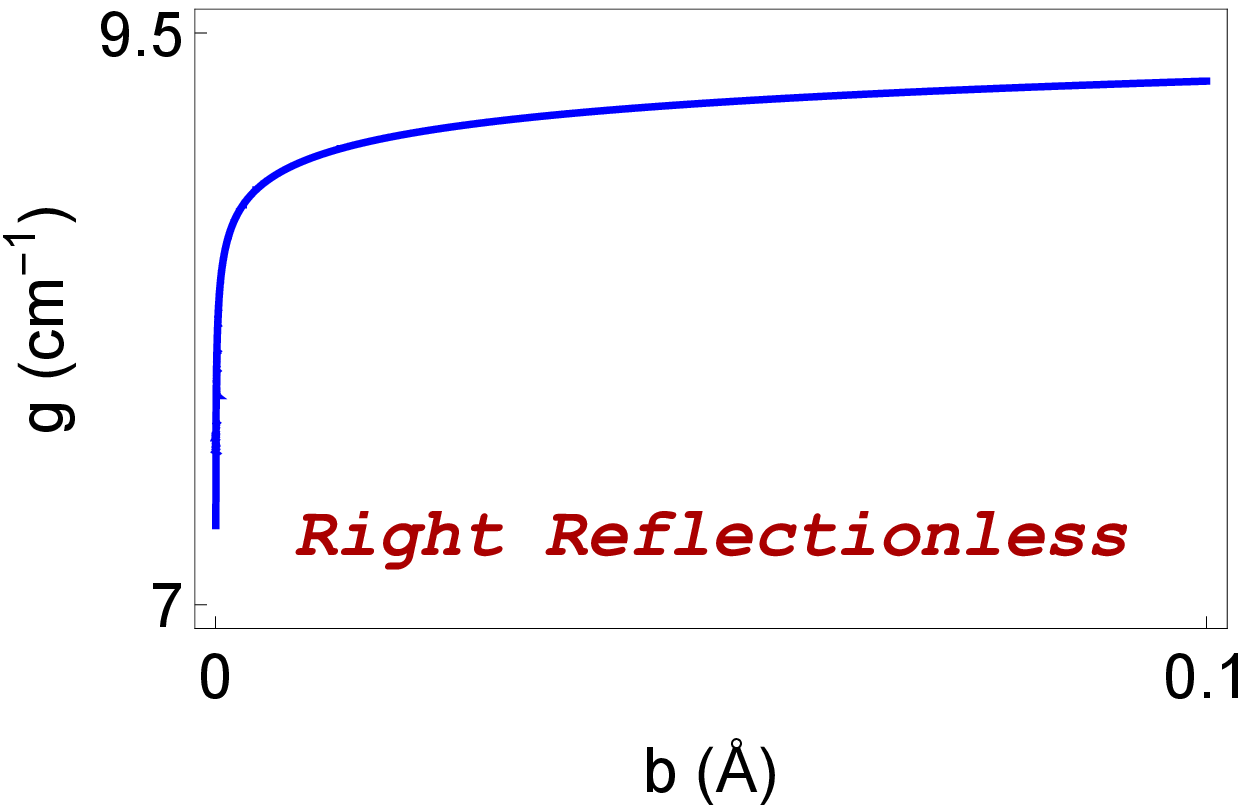}
    \caption{(Color online) Dependence of the gain upon the parameter $b$ to emanate the left and right
reflectionless situations in the presence of 2D Weyl SM.}
    \label{fig2m}
    \end{center}
    \end{figure}

Figure~\ref{fig2m} shows the effect of the parameter $b$ on the gain value in the formation of left and
right reflectionlessness. We see that increasing the value of $b$ results in the left reflectionlessness
at lower gain values whereas it gives rise to the right reflectionlessness at higher gain values.

\begin{figure}
    \begin{center}
    \includegraphics[scale=.32]{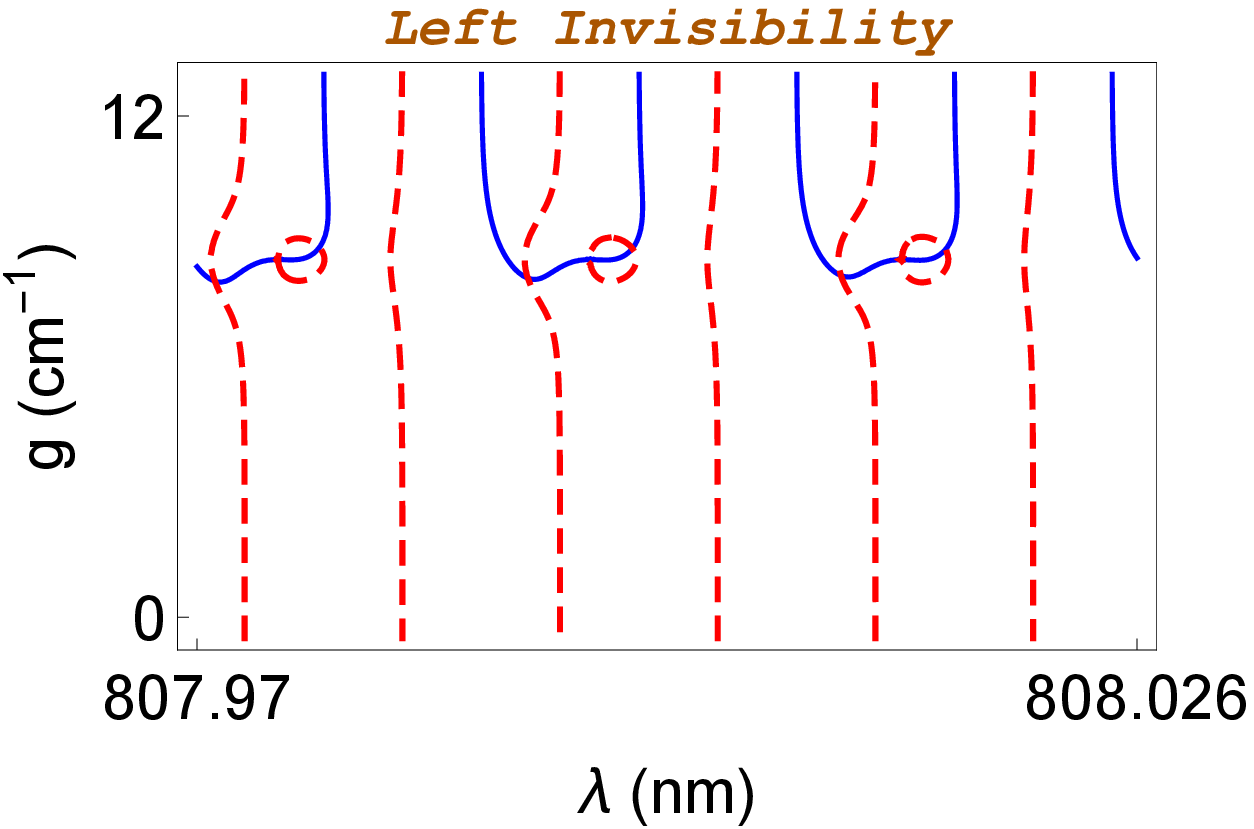}~~~
    \includegraphics[scale=.32]{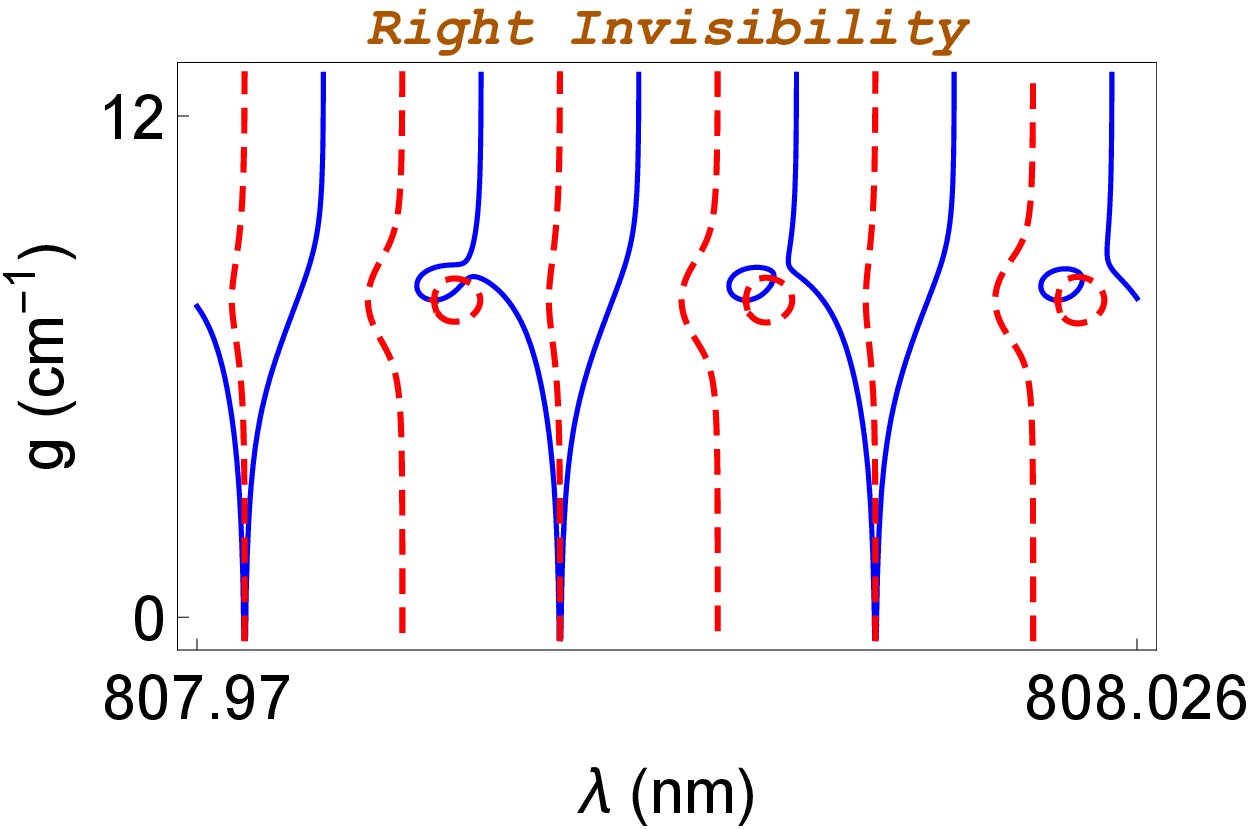}
    \caption{(Color online) Configurations displaying dependence of the left and right invisibility on
the gain coefficient as a function of wavelength in the presence of 2D WSM. Solid blue and dashed red
curves respectively represent the real and imaginary parts of (\ref{invsibilitycondition1}).}
    \label{fig4m}
    \end{center}
    \end{figure}

Invisible configurations require a tedious analysis because (\ref{invsibilitycondition1}) does not
provide a unique expression when decomposed into the real and imaginary parts as opposed to the case of
reflectionlessness. We represent real and imaginary parts separately and consider intersection points as
the corresponding allowed points. Figure~\ref{fig4m} displays the left and right invisibility cases
corresponding to the allowed gain and wavelength values. Here, we distinguish a couple of conclusions.
First one is that left invisibility requires unique and higher gain values at specific wavelength points
while the right invisibility could be obtained at very low and extended gain values at designated
wavelengths. Our second conclusion is that one can not obtain both of the left and right invisibilities
simultaneously with the same gain and wavelength values. This is important because values of the
parameters corresponding to the left and right invisibility address unique values, and one must adjust
the exact parameters to catch the corresponding directional invisibility.

\begin{figure}
    \begin{center}
    \includegraphics[scale=.32]{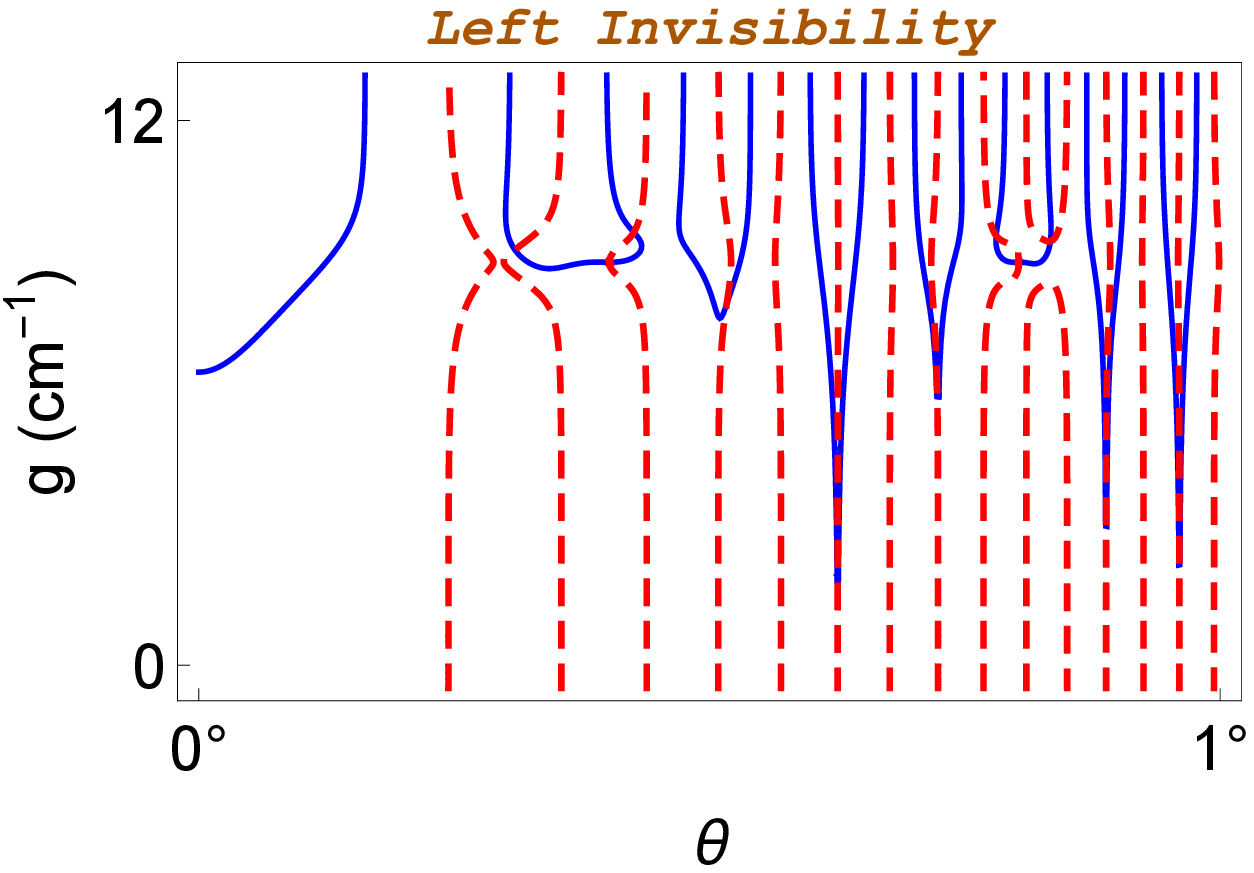}~~~
    \includegraphics[scale=.32]{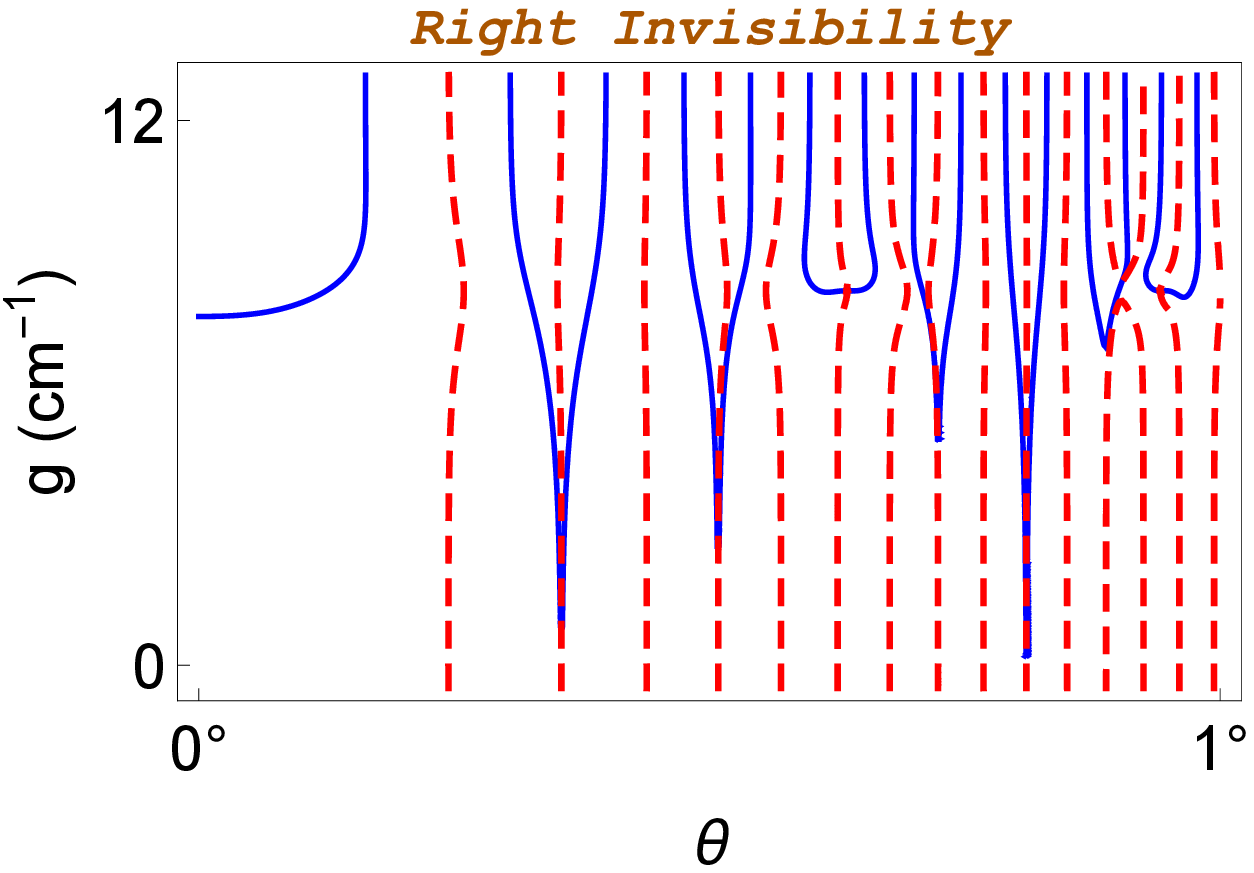}
    \caption{(Color online) Dependence of the gain on the incidence angle for the left and right
invisibility in case of 2D Weyl SM. Again, solid blue and dashed red curves respectively represent the
real and imaginary parts of (\ref{invsibilitycondition1}). }
    \label{fig5m}
    \end{center}
    \end{figure}

Figure~\ref{fig5m} demonstrates the left and right invisibility configurations corresponding to the gain
values as a function of incidence angle $\theta$. Note that not all angles give rise to a unidirectional
invisibility, but only certain angles do it. Left invisibility needs higher gain value compared to the
right one. Likewise, left and right invisibility cases are not simultaneously observed at the same
incidence angles.

\section{Exact Unidirectional Invisibility: Conditions for Broadband and Wide-angle}

Parameters leading to prescribed reflectionless and invisible potentials guide to construct the associated
unidirectional reflectionlessness and invisibility exactly. This is rather simple and smooth mechanism
that gives rise to achieve the aimed consequences through the transfer matrix formalism. In this section,
we rely on the optimal values of the parameters which give rise to desired wide-angle and broadband
invisibility. System parameters dominating the formation of unidirectional invisibility consist of
refractive index $\eta$ which leads to determine the type of slab material to be concealed, slab thickness
$L$, incidence angle $\theta$, wavelength $\lambda$, gain coefficient $g$, and the separation value $b$ of
the Weyl nodes. The role of each parameter was exhibited through the relations (\ref{unidir-refl}) and
(\ref{invsibilitycondition1}). Here, in the light of consequences we obtained, we carry out their effects
in realizing the wide-angle and broadband invisibility case. First of all, we observe that imposing the
condition $\textrm{T} = 1$ into the first two expressions in (\ref{leftreflcoef}) gives rise to
   \be
   \mathbb{R}^{\alpha} = -\alpha\,\zeta_{\alpha}\,\fa_{-\alpha},\label{leftrightreflcoef}
   \ee
where $\zeta_{\alpha}$ is identified by (\ref{zeta}). Here $\alpha$ represents $\pm$, and denotes the left
and right directions respectively. Thus, the left and right invisible configurations are obtained by the
condition $\mathbb{R}^{\alpha} = 0$ once $\mathbb{R}^{-\alpha} \neq 0$ is satisfied. Hence, we should see
the behaviors of the parameters in hand to analyze the quantities $\mathbb{R}^{\alpha}$. We show the
impacts of the effective parameters in plots of $\left|\mathbb{R}^{l}\right|^2$ and
$\left|\mathbb{R}^{r}\right|^2$. We employ the optical construction based on Nd:YAG crystals for the slab
with $\eta = 1.8217$, $L = 1~\mu\textrm{m}$, gain value $g = 8~\textrm{cm}^{-1}$, $\theta = 30^{\circ}$
and $b = 0.05$ {\AA}.

\begin{figure}
    \begin{center}
    \includegraphics[scale=.45]{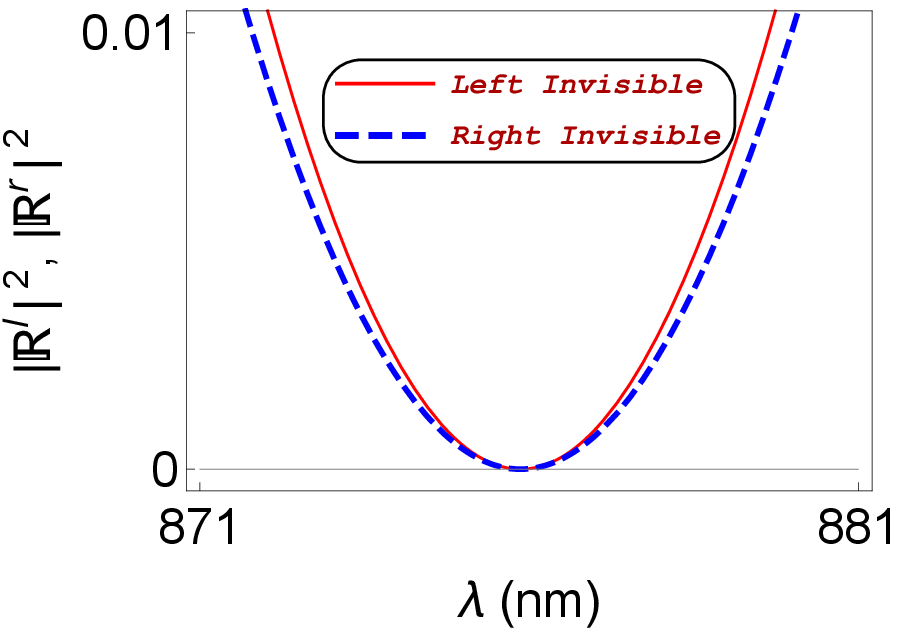}~~~
    \includegraphics[scale=.45]{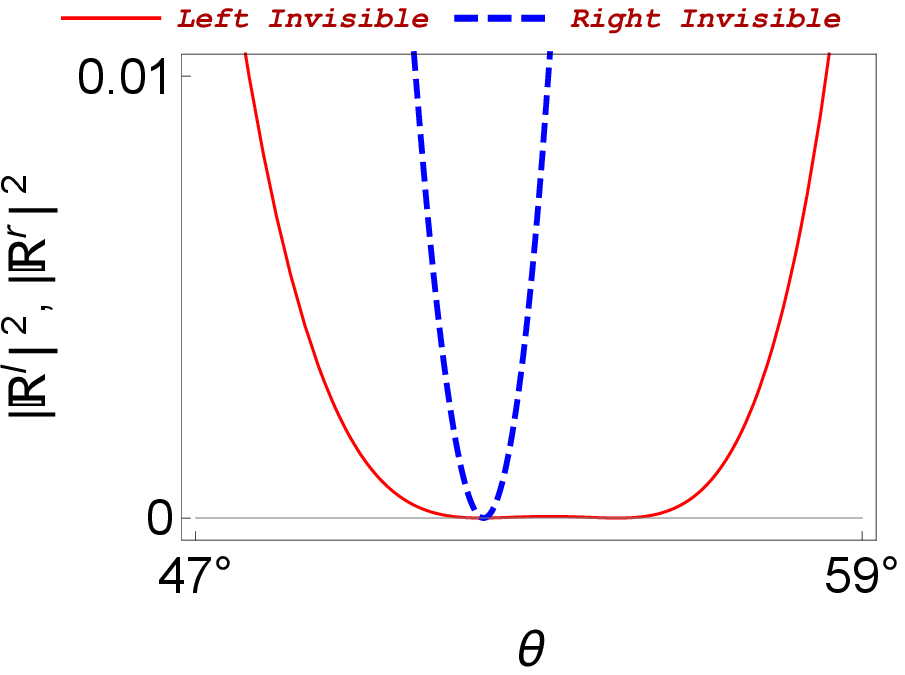}
    \caption{(Color online)  Broadband (left panel) and wide-angle (right panel) invisibilities
corresponding to $1\%$ of flexibility for the case of 2D Weyl SM.}
    \label{broadband}
    \end{center}
    \end{figure}

In Fig.~\ref{broadband}, we present the broadband (left panel) and wide-angle (right panel) invisibility
situations within $1\%$ precision. For the broadband case (left panel), right invisibility bandwidth is
about $\Delta\lambda \approx 8.4~\textrm{nm}$ ranging from $\lambda = 871.7~\textrm{nm}$ to
$\lambda = 880.1~\textrm{nm}$ whereas left invisibility corresponds to the bandwidth
$\Delta\lambda \approx 7.7~\textrm{nm}$ ranging from $\lambda = 872~\textrm{nm}$ to
$\lambda = 879.7~\textrm{nm}$, which is narrower than the right one. As the precision value increases,
broadband invisibility ranges for both left and right directions decrease and approach to each other,
leading to a bidirectional invisibility. As for the wide-angle invisibility case, we notice that left
invisibility angle range is wider than the right one, at which $\Delta\theta \approx 10.5^{\circ}$ for
the left invisibility case corresponding to the range $(47.9^{\circ}, 58.4^{\circ})$ while it is
relatively low for the right invisibility case at which $\Delta\theta \approx 2.5^{\circ}$ corresponding
to the range $(50.9^{\circ}, 53.4^{\circ})$. These broadband and wide-angle invisibility conditions for
a regular material type are rather impressive compared to the pure $\mathcal{PT}$-symmetric slab system and
the case with graphene, at which cases it is impossible to obtain such higher values, see
\cite{pra-2017a, cpa3}.

\begin{figure}
    \begin{center}
    \includegraphics[scale=.45]{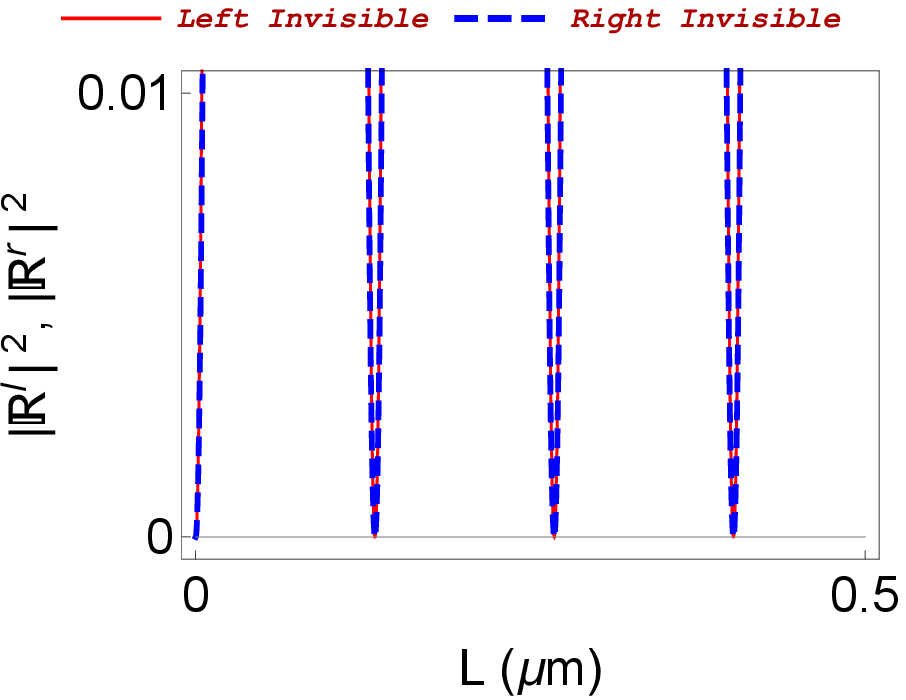}~~~
    \includegraphics[scale=.45]{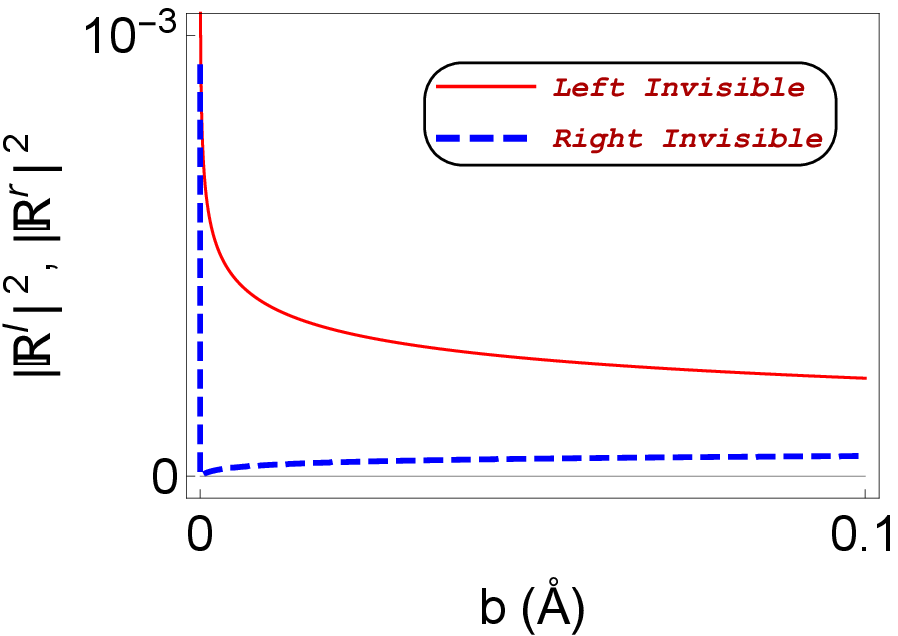}
    \caption{(Color online) The effect of slab thickness (left panel) and parameter $b$ (right panel) on
the broadband and wide-angle invisibility in the case of 2D Weyl SM.}
    \label{broadband2}
    \end{center}
    \end{figure}

We question the effects of slab thickness $L$ (left panel) and separation of Weyl nodes $b$ (right panel)
on the left and right invisibility in Fig.~\ref{broadband2}. It is immediate to see from the left panel
that broadband and wide-angle invisibility can be recovered at finite sizes apart from smaller thickness
values, which is not possible for non-2D WSM cases. This shows that one can achieve broadband and
wide-angle invisibility at larger slab thickness values depending on the adjustment of the remaining
parameters. This is peculiar to just 2D WSM, and normally not expected for other materials cases. Left
panel shows effect of the parameter $b$ such that better results can be obtained with higher $b$ values.

\begin{figure}
    \begin{center}
    \includegraphics[scale=.45]{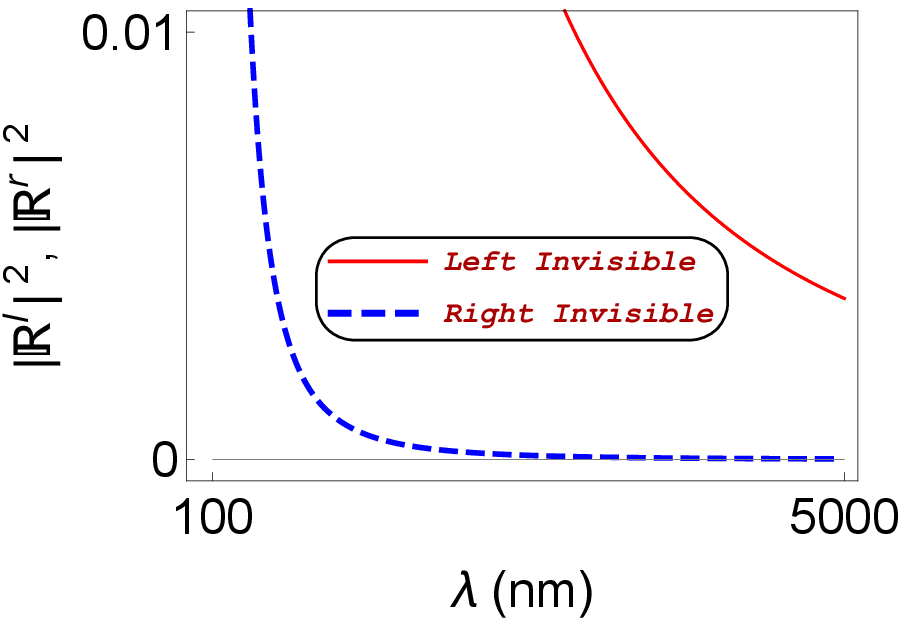}~~~
    \includegraphics[scale=.45]{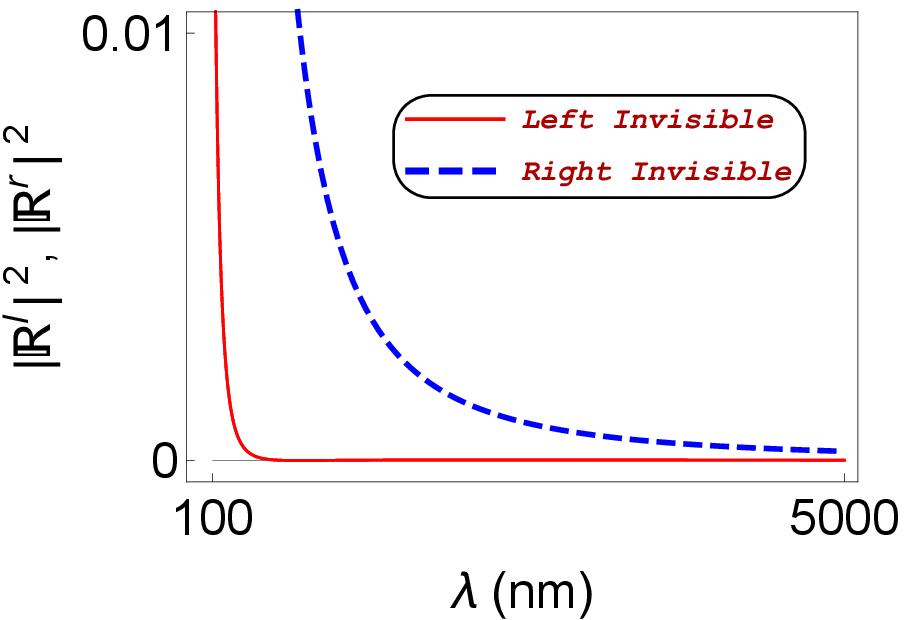}
    \caption{(Color online) Broadband left and right invisibilities corresponding to the incidence angles
of $\theta = 50^{\circ}$ (left panel), and $\theta = 80^{\circ}$ (right panel) within $1\%$ of flexibility
in the case of aluminum slab covered by a 2D Weyl SM.}
    \label{alu1}
    \end{center}
    \end{figure}

\begin{figure}
    \begin{center}
    \includegraphics[scale=.45]{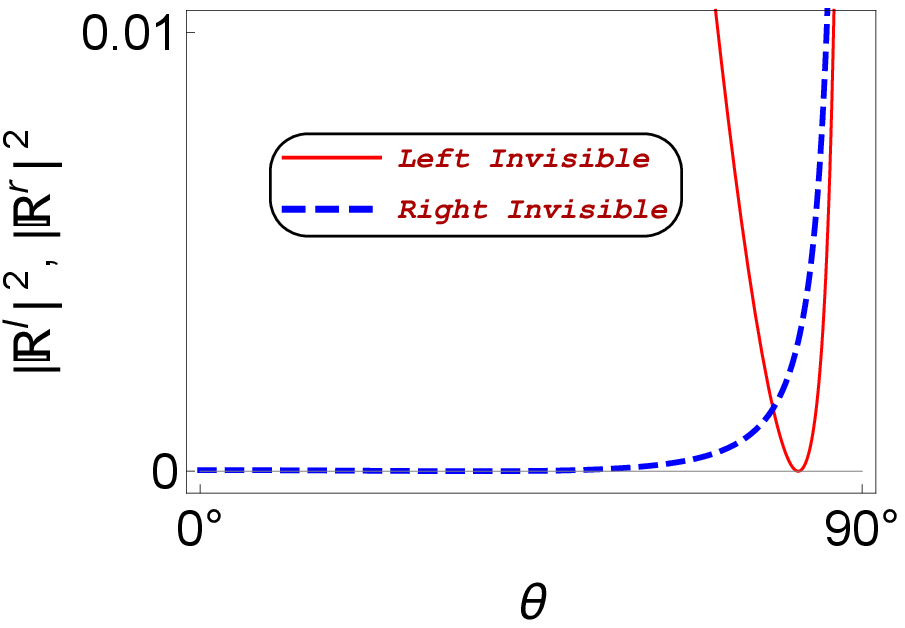}~~~
    \includegraphics[scale=.45]{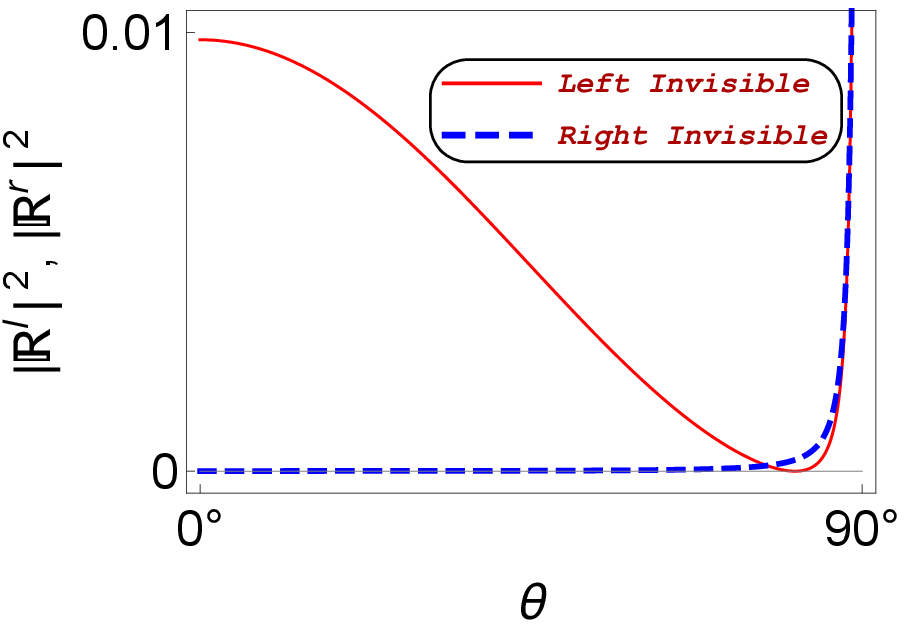}
    \caption{(Color online) Wide-angle left and right invisibilities corresponding to wavelengths
$\lambda = 1500$~nm (left panel) and $\lambda = 5000$~nm (right panel). Slab is made up of aluminum and
surrounded by 2D Weyl SM.}
    \label{alu2}
    \end{center}
    \end{figure}

Furthermore, we investigate the extreme cases to accomplish a much broader band and angle widths by
managing the type of slab material. It is understood that materials whose refractive indices close
$\eta = 1$ display the desired results. For this purpose, we employ aluminum as the slab material whose
refractive index is $\eta_{Al} = 1.0972$,~\cite{alum}. We choose slab thickness as 20~nm, gain
$g= 10$~cm$^{-1}$, and $b$ parameter as 0.05~{\AA}. In Fig.~\ref{alu1}, broadband structure of the left
and right invisibilities are observed for different incidence angles, $\theta = 50^{\circ}$ for the left
panel, while $\theta = 80^{\circ}$ for the right one. We notice that left invisibility is highly
sensitive to the incidence angle. Its broadband coverage extends from $\lambda \approx 100$~nm to whole
low energy regions (Infrared, microwaves and radiowaves) at incidence angle of $\theta = 80^{\circ}$
while it is ranging from $\lambda \approx 2800$~nm at $\theta = 50^{\circ}$. However, right invisibility
is distinguished from $\lambda \approx 750$~nm at angle of $\theta = 80^{\circ}$, and
$\lambda \approx 365$~nm at angle of $\theta = 50^{\circ}$. This shows that small incidence angles favor
the right invisibility whereas large angles favor the left one. This can be observed in Fig.~\ref{alu2}.

In Fig.~\ref{alu2}, the left panel corresponds to $\lambda=1500$~nm and the right one to $\lambda=5000$~nm.
We see that as the wavelength decreases swiftly, the angular range of left invisibility gets lowered
considerably, and described angular range is restricted to above $\theta \geq 60^{\circ}$. Once the
wavelength increases smoothly, wide-angle invisibility range also increases (at high enough wavelength,
one gets all angular range up to $\theta \approx 88^{\circ}$ as the left invisible region). We notice that
right invisibility is easily achieved almost all spectral range at almost all angular ranges (usually up
to $\theta \approx 88^{\circ}$). Thus, one concludes that broadband and wide-angle invisibility is
achieved more flexibly for the right invisibility. However one should regard the correct wavelength and
angular ranges for the left invisibility.

\begin{figure}
    \begin{center}
    \includegraphics[scale=.45]{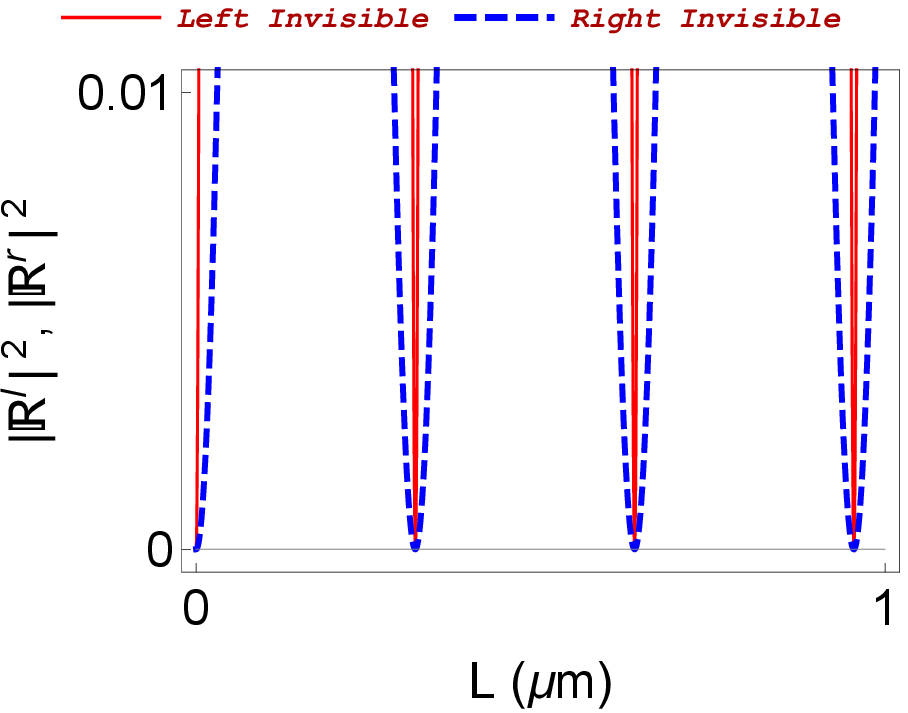}~~~
    \includegraphics[scale=.45]{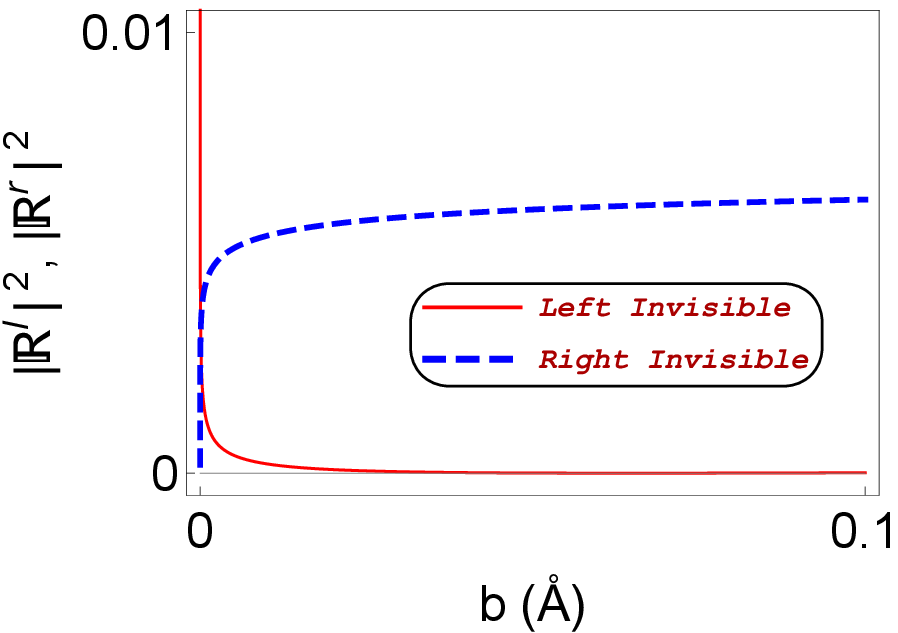}
    \caption{(Color online) The effect of slab thickness (left panel) and parameter $b$ (right panel) on
the broadband and wide-angle invisibility in the case of aluminum slab covered by 2D Weyl SM.}
    \label{alu3}
    \end{center}
    \end{figure}

Figure~\ref{alu3} demonstrates the effect of slab thickness $L$ (left panel) and Weyl node separation $b$
(right panel) on the broadband and wide-angle unidirectional invisibility. We see that even finite values
of slab thickness give rise to desired consequences. This is rather characteristics of the 2D WSM which
is not observed in the case of other material types. The right panel explicitly points out that it is
easier to establish left invisibility compared to the right one at usual Weyl node separations. When the
Weyl nodes are very close, or very small $b$ values, then pure right invisibility is observed, without
allowing the left invisibility.

\section{Concluding Remarks}\label{S9}

In this study we draw a general framework towards understanding the broadband and wide-angle invisibility
phenomenon of $\mathcal{PT}$-symmetric optical slabs with the help of $\mathcal{PT}$-symmetric 2D materials having
scalar conductivity. In particular, we focus our attention to 2D Weyl SMs such that its presence gives
rise to amazing outcomes. Overall optical system is endowed with the property of $\mathcal{PT}$ symmetry to
ensure that essential parameters could be set up properly. No matter which 2D material is placed, we
show that the $\mathcal{PT}$-symmetric slab system results in counter-flowing currents on the surfaces of 2D
materials.

Although our expressions holds for all 2D materials with scalar conductivity, in particular we employed
2D WSMs in our analysis because of their distinct properties. We point out that the effect of 2D WSM
appears through the function $\fu_{\alpha}^{(1,2)}$, which is characterized by its scalar conductivity
$\sigma^{(w)}$, in the components of transfer matrix, see Eq.~\ref{M22=x}. In view of the reflection and
transmission amplitudes, we obtained relations for the exact unidirectional reflectionless and invisible
configurations in (\ref{unidir-refl}) and (\ref{invsibilitycondition1}), which relate the necessary
parameters of the $\mathcal{PT}$-symmetric optical system covered by the $\mathcal{PT}$-symmetric 2D WSM. We
demonstrated the optimal conditions of the parameters contributing to the screening of unidirectional
reflectionlessness and invisibility.

We find out the required gain values, slab thickness, and distance between the Weyl nodes for the
unidirectional broadband and wide-angle invisibility. Notice that the same gain values never give rise
to simultaneous construct of the left and right invisibilities, as opposed to the case of
graphene~\cite{cpa3}. In the case of graphene, broadband and wide-angle structures are rather restricted
in realistic slab materials unless the refractive index of them approaches $\eta = 1$. Desired phenomenon
of unidirectional broadband and wide-angle invisibility is realized at considerably high thickness values.
We observe that the Weyl nodes separation $b$ plays a significant role in the broadband and wide-angle
invisibility, that favors as large $b$ values as possible. We notice that broadband and wide-angle
invisibility is very sensitive to the material type through the refractive index. We performed an extensive analysis of two different kinds of material types, Nd:YAG and aluminum. We show that 2D WSM give
rise to the broadband and wide-angle invisible configurations even with the use of usual materials with
any refractive index, see Fig.~\ref{broadband}. Moreover, when we use a material with a refractive index
close to $\eta = 1$, almost full range broadband and wide-angle invisibility is obtained as is seen in
Figs.~\ref{alu1} and \ref{alu2}. This is quite intriguing and occurs only in the presence of 2D WSM. We
expect our results to guide experimental attempts for the realization of broadband and wide-angle
invisibility with $\mathcal{PT}$-symmetric 2D materials.\\[6pt]

\end{document}